%% file: main.tex
\documentclass[letterpaper, conference]{IEEEtran}
\IEEEoverridecommandlockouts
\usepackage{tablefootnote} % for table footnotes
\usepackage{threeparttable}
\usepackage{cite}
\usepackage{svg}
\usepackage{amsmath}
\usepackage{cuted}
\usepackage{amssymb,amsfonts, epsfig,latexsym}
\usepackage{algorithm}
\usepackage{algpseudocode}

\usepackage{subfigure}
\usepackage{flushend}
\usepackage{soul}
\usepackage[font=scriptsize]{caption} %footnotesize
\usepackage{hhline}
\usepackage{colortbl}
\usepackage{xcolor}
\usepackage{pdfpages}
\usepackage{graphicx, wrapfig}
\usepackage{textcomp}

\usepackage{multirow}
\usepackage{bigstrut}
\usepackage{array}
\usepackage{comment}
\usepackage{booktabs}
\usepackage{multicol}
\usepackage[utf8]{inputenc}
\renewcommand{\arraystretch}{1.5}

\def\BibTeX{{\rm B\kern-.05em{\sc i\kern-.025em b}\kern-.08em
    T\kern-.1667em\lower.7ex\hbox{E}\kern-.125emX}}

\usepackage[none]{hyphenat}
\usepackage[hyphens]{url}

\usepackage[textsize=tiny,textwidth=0.72in]{todonotes}
\usepackage{color} % color added for editing
\usepackage{soul}

\begin{document}

%\title{Custom HLS and HDL Design of LSTM Accelerator for High-Rate Dynamic Applications\\}
%\title{Accelerating High-Rate Dynamic Systems with LSTM Networks\\}
% JDB
%\title{Low Latency Attention Block of Transformer Encoder on FPGA\\}
%\title{FLAME: Flexible Accelerator for the Attention Mechanism of Transformer on FPGA\\}
%\title{FLARE: Flexible Accelerator for the Attention Mechanism in Transformers on FPGA\\}
%\title{FLOAT: Flexible Accelerator for the Attention Mechanism in Transformers on FPGA\\}
%\title{FORAM: Flexible Accelerator for the Attention Mechanism in Transformers on FPGA\\}
%\title{FAMe: Flexible Accelerator for the Attention Mechanism in Transformers on FPGA\\}
\title{FAMOUS: Flexible Accelerator for the Attention Mechanism of Transformer on UltraScale+ FPGAs} %\vspace{-0.5em}
%\author{\IEEEauthorblockN{Ehsan Kabir$^{\mathrm{*}}$, David Andrews$^{\mathrm{*}}$, Miaoqing Huang$^{\mathrm{*}}$}

%\IEEEauthorblockA{\textit{$^{\mathrm{*}}$Department of Computer Science and Computer Engineering, University of Arkansas, USA} \\  
%\{ekabir, dandrews, mqhuang\}@uark.edu}{\centering}}

%\author{\IEEEauthorblockN{ \\ }%Ehsan Kabir$^{\mathrm{*}}$, David Andrews$^{\mathrm{*}}$, Miaoqing Huang$^{\mathrm{*}}$}
%\IEEEauthorblockA{ \\ }}%\textit{}%$^{\mathrm{*}}$Department of Computer Science and Computer Engineering, University of Arkansas, USA} \\  
%\ { \\ }{\centering}
%ekabir@uark.edu, dncoble@email.sc.edu, jsatme@email.sc.edu, austindowney@sc.edu, jbakos@cse.sc.edu,\\ dandrews@uark.edu, mqhuang@uark.edu}}{\centering}

\author{\IEEEauthorblockN{Ehsan Kabir$^{\mathrm{*}}$, Md. Arafat Kabir$^{\mathrm{*}}$, Austin R.J. Downey$^{\mathrm{\$}}$, Jason D. Bakos$^{\mathrm{\dag}}$, David Andrews$^{\mathrm{*}}$, Miaoqing Huang$^{\mathrm{*}}$}
\IEEEauthorblockA{\textit{$^{\mathrm{*}}$Department of EECS, University of Arkansas, Fayetteville},
\textit{Department of $^{\mathrm{\dag}}$CSE, $^{\mathrm{\$}}$ME, University of South Carolina, USA}\\ 
%\textit{$^{\mathrm{\$}}$Department of ME, University of South Carolina, USA}}}{\centering}\\
%\ { \\ }{\centering}
\{ekabir, makabir, dandrews, mqhuang\}@uark.edu, austindowney@sc.edu, jbakos@cse.sc.edu }}{\centering}
%\vspace{-0.8cm} Electrical Engineering and Computer Science, Computer Science and Engineering
%\vspace{-0.8cm}
%\author{\IEEEauthorblockN{}

%\IEEEauthorblockA{\textit{} \\
%\textit{}\\
%\textit{}\\
%}}{\centering}

%\and
%\and
%\and
%\and
%\and

\onecolumn
{\large\vspace*{\fill}

© 2024 IEEE.  Personal use of this material is permitted.  
Permission from IEEE must be obtained for all other uses, in any current or future media, including reprinting/republishing this material for advertising or promotional purposes, creating new collective works, for resale or redistribution to servers or lists, or reuse of any copyrighted component of this work in other works. \\

This work has been accepted as a poster at the FPT 2024 (International Conference on Field Programmable Technology) Proceedings. It will appear in the proceedings as a two-page manuscript and on the IEEE website soon.

% Create vertical space to center the paragraph
\vspace*{\fill}
}
\twocolumn

\maketitle

\input{abstract}

\begin{IEEEkeywords}
FPGA, Transformer, Attention, High-Level Synthesis, Natural Language Processing, Accelerators.%, Custom hardware.
\vspace{-0.2cm}
\end{IEEEkeywords}

\input{introduction}

\input{background}
%\input{dynamics}
%\input{model}
%\input{relatedwork}
\input{architecture}

%\input{hls}
%\input{hdl}
%\input{system}
%\input{implementations}
\input{evaluation}

%\input{analytical}
\input{conclusion}
\small
\bibliographystyle{IEEEtran}
%\vspace{-0.2cm}
\bibliography{IEEEabrv,tnn, tnn_base}  

%\vspace{12pt}
%\color{red}
%\input{reviews}
%\input{rebuttal}

\end{document}

%% file: abstract.tex
\begin{abstract}
%%Transformer neural networks (TNNs) are being applied across a widening range of application domains, including natural language processing (NLP), machine translation, and computer vision (CV). Their popularity is largely attributed to the exceptional performance of their multi-head self-attention blocks when analyzing sequential data and extracting features. To date, there are limited hardware accelerators tailored for this mechanism, which is the first step before designing an accelerator for a complete model. 

This paper proposes \textit{FAMOUS}, a flexible hardware accelerator for dense multi-head attention (MHA) computation of Transformer neural networks (TNNs) on field-programmable gate arrays (FPGAs). It is optimized for high utilization of processing elements and on-chip memories to improve parallelism and reduce latency. An efficient tiling of large matrices has been employed to distribute memory and computing resources across different modules on various FPGA platforms. The design is evaluated on Xilinx Alveo U55C data center cards containing Ultrascale+ FPGAs. Experimental results showed that it can attain a maximum throughput, the number of parallel attention heads, embedding dimension, and tile size of 328 (giga operations/second (GOPS)), 8, 768 and 64 respectively on the U55C. Furthermore, it is 3.28$\times$ and  2.6$\times$ faster than the Intel Xeon Gold 5220R CPU and NVIDIA V100 GPU respectively. It is also 1.3$\times$ faster than the fastest state-of-the-art FPGA-based accelerator.
% which CPU and GPU ? and U200 
% and the proposed architecture achieves ...×, ....× improvement in latency, and ...×, ...× energy savings compared to CPU and GPU. and 1.29 $\times$

\end{abstract}

%% file: introduction.tex
\section{Introduction}
Transformer neural networks have demonstrated significant advancements in natural language processing (NLP), machine translation, computer vision \cite{attention, wang_via_2022}, and other domains in recent years. %%Numerous transformer-based models have surfaced, including full transformers containing an encoder and decoder \cite{attention}, BERT \cite{PretrainedTF, bert}, Transformer-XL \cite{TXL}, ALBERT \cite{albert}, T5 \cite{t5}, Routing Transformers \cite{routing}, structBERT \cite{structbert}, and more. These models  %% language, \cite{attention}, \cite{NMT}
They contain a remarkable feature named multi-headed attention (MHA) mechanism which is different from the traditional convolutional neural network (CNN), recurrent neural network (RNN), and long short term memory (LSTM) model.
%It is even replacing RNNs and LSTMs for NLP tasks, as well as convolutional layers in CV tasks, because
It enables a high level of computational parallelism for both the training and inference phases making it highly suitable for acceleration on hardware like GPUs and FPGAs, with FPGAs being particularly advantageous due to their high degree of parallelism, low latency, and energy efficiency \cite{dl}.  %% ,gan_hardware-aware_2022 %%Nevertheless, the attention mechanism incurs high computational costs due to intensive matrix computations and intricate data flow \cite{sanger, ham_elsa_2021}. %%It consumes a significant amount of runtime in many existing TNNs, ranging from about 38\% to 64\% when the sequence length (number of tokens in the input sequence) varies from 64 to 256\cite{ham_elsa_2021}. 
%%Unfortunately, general-purpose platforms such as GPUs and CPUs are inefficient for processing TNNs due to low computational efficiency, underutilized memory bandwidth, and significant compilation overheads\cite{flightLLM}. In contrast, 
%%FPGAs have gained widespread use for accelerating DNNs due to their high level of parallelism and low latency \cite{dl, gan_hardware-aware_2022}. %%Many works focus on parallelizing computations to accelerate CNN, LSTM, Graph Convolutional Network (GCN) \cite{cnn_low_batch, lstm_high_rate, gcn_large, que_optimizing_2022} on FPGAs. Recently, Some works have successfully built FPGA or application-specific integrated circuit (ASIC) hardware accelerators for transformers \cite{lu_hardware_2020, A3AA, peng_length_2022}.
Most of the FPGA or ASIC-based hardware accelerators for transformers \cite{peng_accelerating_2021} %compress the model by using different weight pruning strategies to accelerate attention, and they reduce latency by incorporating sparse matrices. Thus, they %lu_hardware_2020, A3AA,
have specialized sparse architecture for a specific application. %%However, different applications require different sparsity patterns, necessitating the redesign of hardware architectures for optimal results, which is a time-consuming and challenging task.
Thus, they lack the flexibility to be reconfigured for a different model during runtime.
%% ASICs are designed for a specific model and configuration, so, they perform poorly on different models or even the same model with varying configurations\cite{FaFlA}. Custom FPGA accelerators also lack the flexibility to be reconfigured for a different model during runtime. %%Thus, a versatile accelerator is needed that can efficiently handle dense matrix computations across various TNN applications. 
%%The study in \cite{lu_hardware_2020} utilizes logic resources to implement a systolic array (SA) for parallelism, leading to a waste of digital signal processing (DSP) resources that are capable of high-speed computation at higher frequencies. DSP consumption also depends on the implementation method. For example, 
Most works \cite{peng_accelerating_2021} used high-level synthesis (HLS) tools, but it is challenging to write efficient HLS code that can effectively manage certain FPGA resources like DSPs for optimal performance\cite{lstm_high_rate}. %% , jiang_ultra_nodate wojcicki_accelerating_2022 %% while some used hardware description language (HDL) \cite{chen_high-frequency_2023, yang_efa-trans_2022, bai_ltrans-opu_nodate} for design. %This tool can automatically compile the C/C++ code into a netlist to implement different functions.While HLS takes less implementation time compared to HDL, it is challenging to write efficient HLS code that can effectively manage certain FPGA resources like DSPs on an FPGA for optimal performance\cite{lstm_high_rate}.  
%%Analysis done in \cite{mha_compute, li_ftrans_2020, luo_calabash_2023} showed that 
Furthermore, MHA uses a large amount of the block RAMs (BRAM) \cite{luo_calabash_2023}. Since FPGAs usually have limited BRAM, creating a good partitioning scheme that works well with the architecture is necessary and can be challenging. %%mha_compute, li_ftrans_2020, %%input matrices must be partitioned into tiles. However, formulating an ideal partition scheme that aligns well with the architecture poses a considerable challenge.  %%and has the highest number of operations. 

%%In this paper, we present \textbf{\textit{FAMOUS}}, a flexible accelerator designed to adapt to a wide range of TNN applications. Our HLS-based code is optimized to utilize more DSPs and BRAMs in parallel. %The computations are performed using DSP48 slices, while data for the DSPs is stored in BRAMs or registers within the processing elements (PE). 
%%\textbf{\textit{FAMOUS}} integrates efficient tiling along with enhanced parallel computation and communication to accelerate the attention mechanism as much as possible.
%In this paper, we designed a flexible accelerator called \textbf{\textit{FAMOUS}} which is versatile enough for various TNN applications. Our code written in the HLS tool is optimized enough to utilize more DSPs and BRAMs in parallel. The computations are done by DSP48 slices, and the data used by the DSPs is stored in BRAMs or registers inside the processing elements (PE). 
%\hfill \\
%\hfill \\

\textit{To address these challenges, this paper makes the following contributions:}%The contributions of this paper are:}
%\vspace{-3.5mm}
\begin{itemize}
    %\item [$\bullet$] An accelerator for the attention mechanism of transformer that achieves low latency by exploiting the available parallelism of FPGA.  
      
    \item [$\bullet$] An efficient tiling of weight matrices to accommodate large models in on-chip memory.
    
    \item [$\bullet$] A novel architecture ensuring high BRAM and DSP utilization for efficient parallel processing of the transformer's attention mechanism with low latency.%%A novel architecture that ensures high utilization of BRAM and DSP, enhancing parallel processing of the attention mechanism of the transformer and achieving low latency..
    
    \item [$\bullet$] A parameterized HLS code that enables users %ensures the design time modification capability of 
    to modify some parameters at design time from HLS tool. %such as the number of attention heads, the tile size, the embedding dimension, the bit width, and the sequence length 
    
    \item [$\bullet$] A runtime programmable feature that enables users to modify some parameters at runtime from software.  

    %\item [$\bullet$] A theoretical model to validate both predicted and experimental latency.

    %\item [$\bullet$] A runtime programmable feature of the parameters such as attention heads, embedding dimension, and sequence length to evaluate different topologies without re-synthesizing the hardware.  
    
\end{itemize}

%% file: background.tex
\section{Background}\label{back}
\begin{figure}[h]
\centering
    %\begin{subfigure}{.5\textwidth}
    \centering
    \includegraphics[height=3.0cm, width=0.6\linewidth]{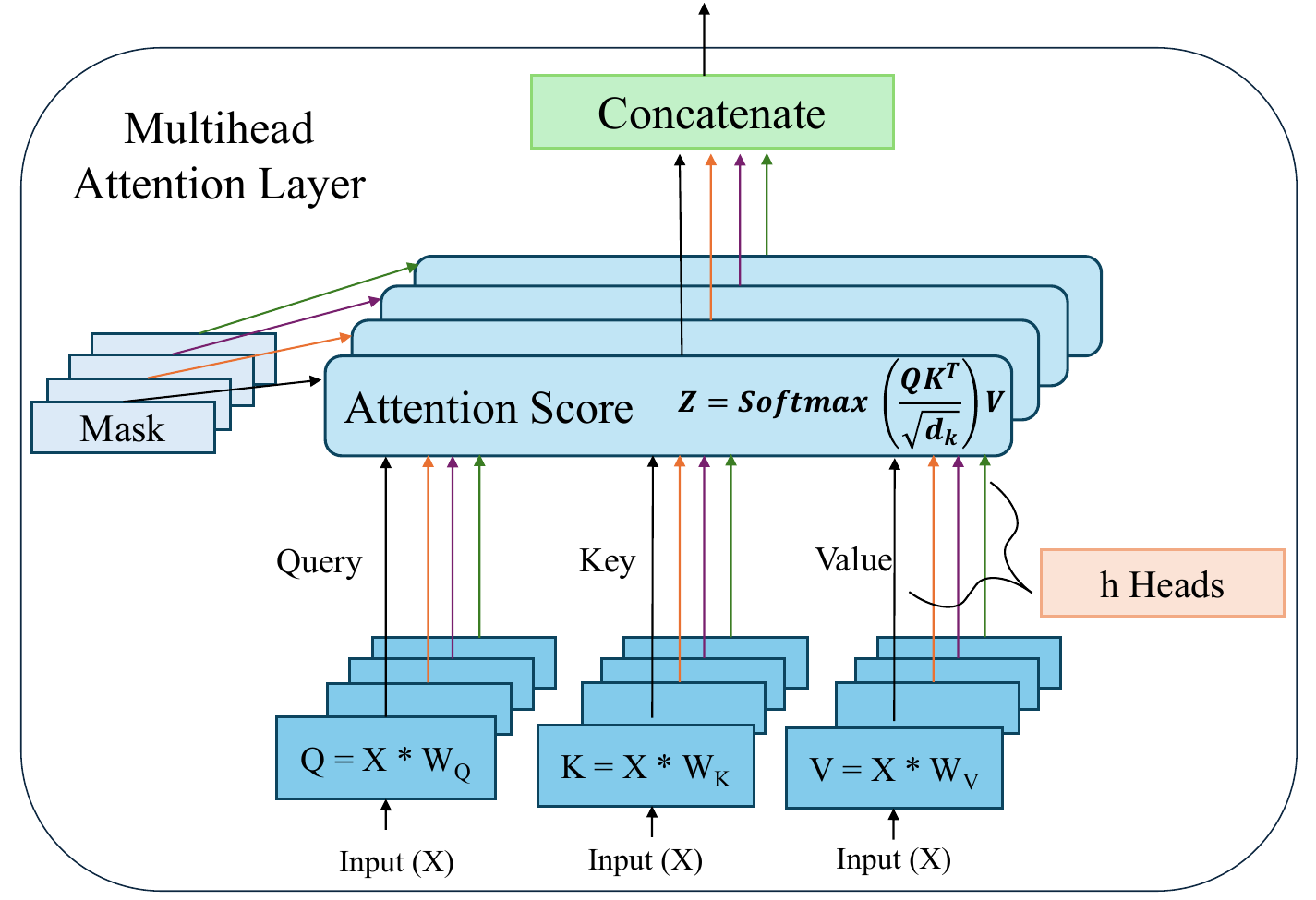}
    \caption{\label{mha}Multihead Attention Layer.}
\end{figure}

There are several building blocks in transformers of which the multi-head attention (MHA) is described here. Fig.~\ref{mha} illustrates the scaled dot product attention in each head, which is a crucial part of the MHA layer. %%The attention weights are computed by performing the dot product of the Q and K matrices and subsequently scaling them down by the square root of the 2\textsuperscript{nd} dimension of the K matrix. This scaling is essential to prevent the dot products from becoming excessively large, contributing to the stabilization of gradients during the training process. Subsequently, the scaled dot products undergo the softmax function, resulting in the computation of attention weights. These weights are then utilized to perform a weighted sum of the value vectors. The ultimate output is the projection of the concatenated sequences from all heads.
The output of MHA can be represented as Equation 1 \& 2. The input sequence X is linearly mapped into $Q_i, K_i, V_i$ matrices using weights and biases. The parameter $d_k = d_{model}/h$ is the 2\textsuperscript{nd} dimension of $Q_i$ and $K_i$. $d_{model}$ is a hyperparameter called embedding dimension and h is number of heads. `i' is the index for attention heads.
\vspace{-0.4cm}
\vspace{-0.4cm}
{\small
\begin{gather}
\label{eq:attention}
\begin{split}
    Attention (Q_i, K_i, V_i) = softmax \left( Mask \left(\frac{Q_iK^T_i}{\sqrt{d_k}} \right)\right)V_i
%\end{equation}
\end{split}\\
%\begin{align}
\begin{split}
    Q_i = X \times W_q + B_q, K_i = X \times W_k + B_k, V_i = X \times W_v + B_v
\end{split}
%\end{align}
\end{gather}}
\vspace{-0.6cm}
%\begin{equation}
    %\resizebox{.8\hsize}{!} 
%    Q_i = X\times W_q + B_q, K_i = X\times W_k+ B_k, V_i = X\times W_v + B_v
%\end{equation}
%\end{comment}

%% file: architecture.tex
\section{Accelerator Architecture}\label{sec2}
\begin{figure}[h!]
\centering
\includegraphics[height=3.0cm, width=0.8\linewidth]{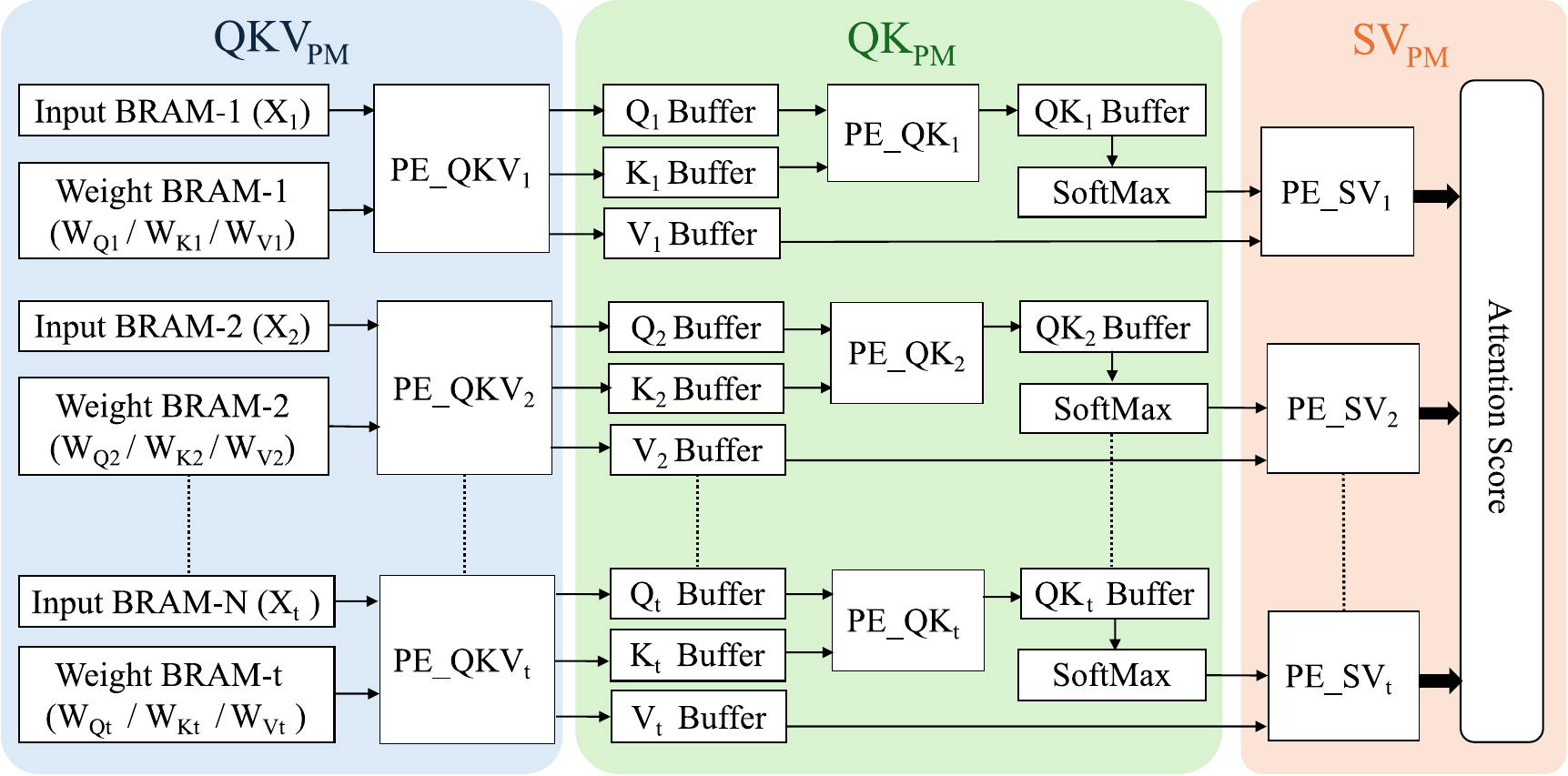} % 6cm 0.7cm
\vspace{-0.2cm}
\caption{\label{mha_System}Accelerator Architecture for Attention Mechanism}
\end{figure}

The core of the accelerator shown in Fig.~\ref{mha_System} was designed in C language on Vitis high-level synthesis (HLS) 2022.2.1 tool. %%Functional verification was performed through its C simulation and C/RTL co-simulation features. This section describes the HLS design technique that generates an optimized architecture utilizing most of the BRAMs and DSPs in the processing elements, ensuring high parallelism.
%%\subsection{Overall Structure} It is shown in Fig.~\ref{mha_System}. 
There are three main processing modules in it. They are denoted as ${QKV}_{PM}$, ${QK}_{PM}$ and ${SV}_{PM}$ according to the output they produce. The number of instances for these modules depends on the number of attention heads (h). Each module contains an array of processing elements (PE). A PE is comprised of a DSP48 performing multiplication and accumulation (MAC) operations. The number of PEs (t) depends on the unrolling factor of the inner loop and the initiation interval of the pipelined outer loop. %%The PE array's data access pattern and computational requirements differ across modules. Therefore, they are defined separately with distinct sets of PE arrays. This approach enables the optimization of each module separately. Input data and weights are stored in multiple BRAMs to enable parallel access.
${QKV}_{PM}$ module generates the query, key, and value matrices. The arrays used in this module are divided into subarrays using our tiling technique to fit into the BRAMs. ${QK}_{PM}$ module performs the matrix-matrix multiplication operations between the Q and K matrices. As these matrices are relatively small, they are not tiled. The output from ${QK}_{PM}$ module is transmitted to the ${SV}_{PM}$ module after softmax operation, where it undergoes matrix-matrix multiplication operations with the value (V) matrix.
%%In our architecture, each PE is independent, with its own local memory, control and computing unit. The weights ($W_q$, $W_k$, $W_v$) for generating query (Q), key (K) and value (V) matrix are declared as separate two-dimensional arrays of size ($\frac{d_{model}}{h} \times TS$) in HLS. 
%%
%%TS is tile$\_$size which represents the dimension of the sub-matrices into which the larger weight matrices are divided. The number of heads and tiles, and the array partition directive on HLS determine how the arrays will be partitioned to generate multiple two-port BRAMs. Due to the limited ports of BRAMs, array partitioning and data loading are efficiently managed to ensure that data required simultaneously by a DSP are stored in separate BRAMs. The Q, K, and V matrices of size ($SL\times \frac{d_{model}}{h}$) are stored in intermediate buffers, which are also implemented as BRAMs. SL is sequence$\_$length.
\vspace{-0.4cm}

\vspace{-0.5cm}
\section{Tiling Technique}
\begin{figure}[h]%{8cm}%{0.8\linewidth}
\centering
%\includegraphics[width=1\linewidth]{figures/arch-n.png}
%\begin{center}
\includegraphics[height=2.5cm, width=1.0\linewidth]{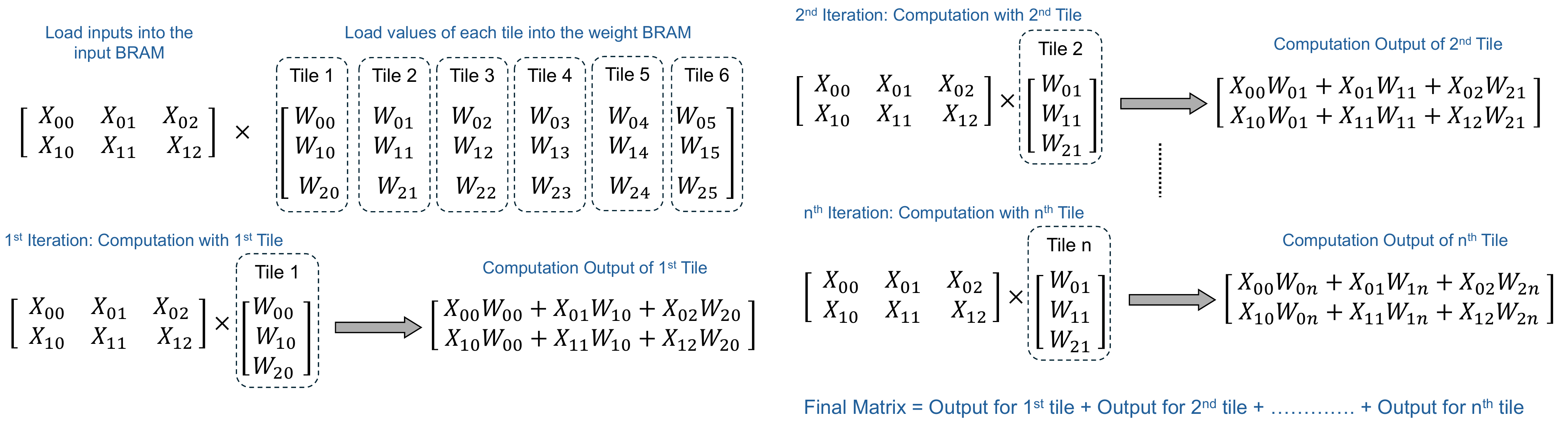} % 6cm 0.7cm
%\end{center}
\vspace{-0.4cm}
\caption{\label{tile_tech_mha}Tiling Technique in Multihead Attention Layer.}
\end{figure}

%%As transformer models tend to be large, tiling helps prevent excessive utilization of on-chip memory and computing units. It also ensures that HLS tool can effectively partition arrays, and pipeline or unroll the loops to minimize latency within a short compilation time. 
Fig.~\ref{tile_tech_mha} describes our unique tiling strategy. The weight matrices are tiled along the second dimension (column of the matrix) only because the first dimension (row of the matrix) is already reduced by the number of heads. %%The weight matrices are partitioned into tiles, allowing BRAMs to be loaded with partial data retrieved from off-chip memory.
\vspace{-0.4cm}
\vspace{-0.2cm}
Thus, they are loaded ($\frac{d_{model}}{TS}$) times. Input buffers of each attention head are declared as a two-dimensional matrix of size (SL $\times$ TS). Therefore, tiling is applied along the column of the matrix, and they are also loaded ($\frac{d_{model}}{TS}$) times. TS is tile$\_$size and SL is sequence$\_$length.

%%which represents the dimension of the sub-matrices into which the larger weight matrices are divided. Therefore, tiling is applied along the column of the matrix, and they are also loaded ($\frac{d_{model}}{TS}$) times. At each iteration, data for only one tile is loaded first. The PEs then perform computations on this data, storing the results in intermediate buffers. These results are also accumulated with those from previous iterations in the next cycle. Consequently, the final output is the cumulative sum of the outputs computed for all the tiles. 

%\vspace{-0.2cm}
\section{Runtime Programmable Feature}
\begin{figure}[h]
\centering
\includegraphics[height=2.5cm, width=0.7\linewidth]{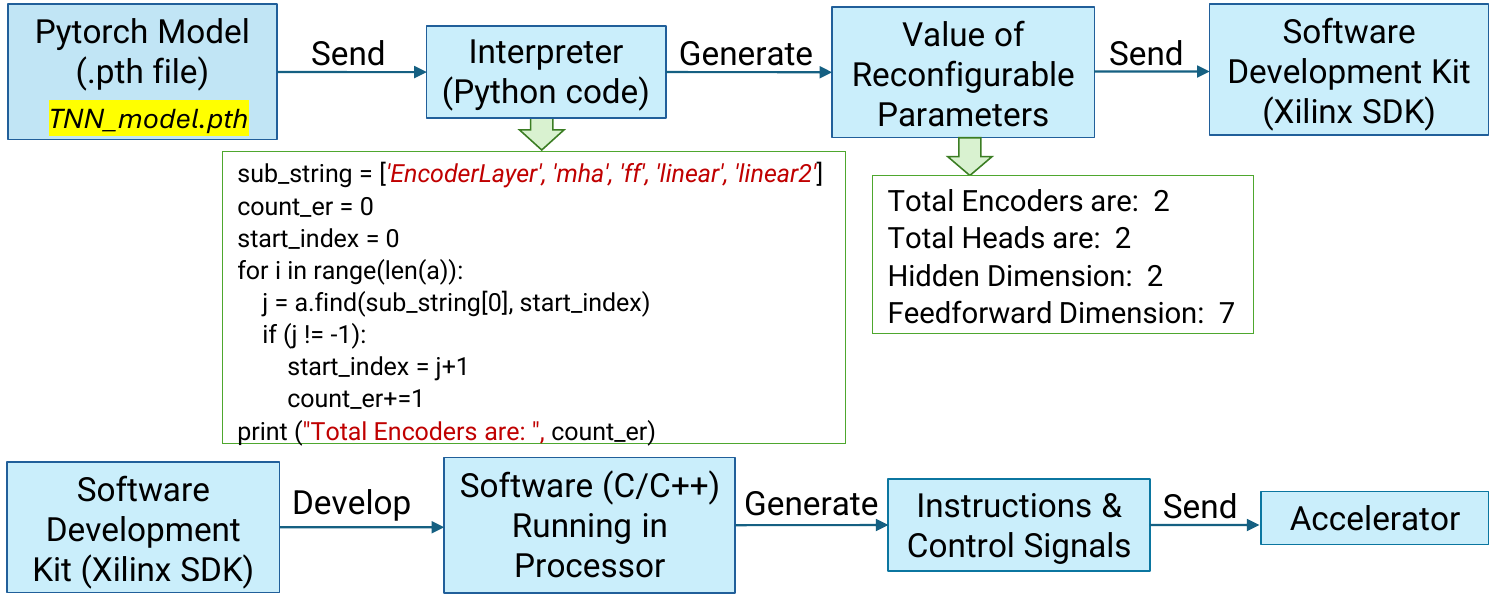} % 6cm 0.7cm
\caption{\label{isa} Process for Incorporating Programmability.}
\end{figure}

The parameters such as attention heads, embedding dimension, and sequence length were runtime programmable. These parameters can be sent to \textbf{\textit{FAMOUS}} from the software using the steps shown in Fig.~\ref{isa}. %TNN models were trained using the PyTorch framework, and the suitable models should be saved as \textit{'.pth'} files. These files were then sent to a Python interpreter to extract the data about attention heads, embedding dimension, and sequence length. These data will differ across applications, but \textbf{\textit{FAMOUS}} will not need resynthesis for each one. Only the software code will need some modifications. The Xilinx SDK tool was used to write the software in C++, which runs on the processor. The extracted data such as the number of attention heads, embedding dimension etc. from the interpreter was used in this software. Based on this data, the processor generated instructions and control signals for the accelerator, allowing it to activate different parts of the hardware.
\vspace{-0.4cm}
\vspace{-0.4cm}
%\hfill \\
%\hfill \\

%\vspace{-0.4cm}
%\vspace{0.1cm}

%
%feature of the to evaluate different topologies without re-synthesizing the hardware.

%% file: evaluation.tex
\section{Evaluation and Results}\label{results}

Table~\ref{results-overall} illustrates the runtime programmable capability, resource utilization, and performance of \textbf{\textit{FAMOUS}}. %%Synthesis was performed once for a constant tile size. The design parameters such as embedding dimension ($d_{model}$), number of heads (h), and sequence length (SL) of the accelerator were configured before synthesis with the fixed values of 768, 8, and 64, respectively, according to a variant of BERT\cite{bert} and the available FPGA resources. Then they were dynamically adjusted during runtime using $\mu$B. Hence, FAMOUS can be synthesized for a fixed number of resources, but it will remain flexible enough to accommodate smaller architectures as needed. The tile size can be adjusted only before synthesis. The data was quantized into 8-bit fixed-point numbers. Quantization for various applications may lead to accuracy loss, although it wasn’t our primary focus. If a larger bit width is necessary, the design can be easily adjusted by modifying certain parameters in HLS code during design time, which will affect resource utilization and latency.
Tests 1, 2, and 3 examine the effect of varying the number of heads, tests 4 and 5 evaluate changes in embedding dimensions, and tests 6, 7, and 8 analyze variations in sequence length, all in relation to latency and throughput (GOPS (giga operations per second)). %%can be varied within the same accelerator dynamically affecting the latency and throughput where throughput is defined as the number of giga operations per second (GOPS).
On Alveo U55C, the lowest latency of 0.94 ms and the highest GOPS of 328 were achieved for 8 parallel heads when the tile size was 64. %Tests no. 4 \& 5 show the effect of varying embedding dimensions on performance where latency increased and GOPS decreased for a larger dimension. Sequence length was dynamically varied for tests no. 6, 7 \& 8, and performance deteriorated as the length increased. It can be observed that resource utilization remained unchanged from tests 1 to 8 because the accelerator was synthesized only once when tile size was constant, while other parameters could be reconfigured at runtime from the software. We ensured high resource utilization levels, with 46\% DSPs, 78\% BRAMs, and 98\% LUTs. The optimal number of attention heads operating in parallel was determined to be 8 when the tile size is 64 on Alveo U55C.
Table~\ref{compare} compared \textbf{\textit{FAMOUS}} with some GPUs and CPUs running approximately at 1.5GHz frequency. We achieved 3.28$\times$, 2.6$\times$, 1.17$\times$ speed up, and an increase in throughput compared to Intel Xeon Gold 5220R CPU, NVIDIA V100 GPU, and Intel E5 2698 v4 CPU respectively because of higher parallelism. %%Topologies include sequence length, embedding dimension, and number of heads. Assuming that attention heads operate in parallel, their number should not impact the latency. Therefore, we did not alter the number of attention heads, even though other works used different numbers. However, we varied the embedding dimensions in line with other studies to ensure a fair comparison. We achieved 3.28$\times$, 2.6$\times$, 1.17$\times$ speed up and %3.28$\times$, 2.6$\times$, 1.17$\times$ increase in throughput compared to Intel Xeon Gold 5220R CPU, NVIDIA V100 GPU, and Intel E5 2698 v4 CPU respectively because of higher parallelism.  %operates at a frequency of 400 MHz and
Table~\ref{fpga} compared \textbf{\textit{FAMOUS}} with other FPGA-based accelerators. %%To ensure a fair comparison, we presented the latency specifically for the computation of the attention mechanism, excluding the latency associated with load and store operations for the accelerator, in this table. 
Our latency is lower and GOPS is higher than all other works except for Calabash\cite{luo_calabash_2023} because it excluded computation time for Q, K, and V calculations. %%Calabash claims to outperform the Intel E5 CPU with a $2.5 - 6.7\times$ speedup. %%The works in \cite{lu_hardware_2020}, \cite{ye_accelerating_2023}, and \cite{li_unified_2023} reported latency for single-head attention calculation.  
%%Their embedding dimensions are also smaller in this table. The number of operations (GOP) reported for the transformer base model in \cite{li_unified_2023} is equivalent to the operations of a single self-attention head. It compared latency with \cite{lu_hardware_2020}, which reported latency for single-head computations based on the given resource usage and algorithm. Ye et al. \cite{ye_accelerating_2023} compared its results with \cite{lu_hardware_2020}, so it was assumed that it also measured latency and resources for single-head computation. Since our results were for 8 self-attention heads, we multiplied the results of other works by 8 for a fair comparison. It also compared with Peng et al. \cite{peng_accelerating_2021} though Peng et al. implemented a complete transformer model. However, it provided latency details for each layer, allowing us to extract the latency specifically for the attention layer.
%%\vspace{-0.2cm}
\begin{table}[h]
\setlength{\arrayrulewidth}{0.05pt}%{0.1mm}
\renewcommand{\arraystretch}{1.0}
  \centering
  \caption{Overall Results for MHA Accelerator.}
  \vspace{-0.2cm}
  \resizebox{1.0\linewidth}{1.5cm}{%3.0cm 0.9cm %c
    \begin{tabular}{|c|c|c|c|c|c|c|c|c|c|c|c|c|}
    \hline
    %\noalign{\hrule height 2pt} !{\vrule width 2pt}
   \multicolumn{1}{|c|}{\multirow{2}[2]{*}{\textbf{Test no.}}} & \textbf{Sequence} & \textbf{Embedding } & \multicolumn{1}{c|}{\textbf{Number}} & \textbf{Tile} & \multirow{2}[2]{*}{\textbf{FPGA}} & \textbf{Data} & \multirow{2}[2]{*}{\textbf{DSPs}} & \textbf{BRAMs} & \multirow{2}[2]{*}{\textbf{LUTs}} & \multirow{2}[2]{*}{\textbf{FFs}}& \textbf{Latency} & \multicolumn{1}{|c|}{\multirow{2}[2]{*}{\textbf{GOPS}}} \bigstrut[t]\\
   %!{\vrule width 2pt} !{\vrule width 2pt}  & \textbf{Frequency} 
   \multicolumn{1}{|c|}{ }   & \textbf{ Length} & \textbf{Dimension} & \multicolumn{1}{c|}{\textbf{of Heads}} & \textbf{Size} &   & \textbf{Format} &   & \textbf{18k} &   &   & \textbf{ (ms)} &  \multicolumn{1}{|c|}{}\bigstrut[b]\\
    %\hline !{\vrule width 2pt} & \textbf{ (MHz)}
    \noalign{\hrule height 2pt}
    \#1 & \multirow{3}[6]{*}{64} & \multirow{3}[6]{*}{768} & \multicolumn{1}{c|}{8} & \multirow{3}[6]{*}{64} &  \multirow{2}[2]{*}{Alveo}  & \multirow{3}[6]{*}{8bit fixed} & \multirow{3}[6]{*}{4157 (46\%)} & \multirow{3}[6]{*}{3148 (78\%)} & \multirow{3}[6]{*}{1284782 (98\%)} & \multirow{3}[6]{*}{661996 (25\%)}  & 0.94 & 328 \bigstrut\\
\cline{1-1}\cline{4-4}\cline{12-13}    \#2 &   &   & \multicolumn{1}{c|}{4} &   & \multirow{2}[2]{*}{U55C} &   &   &   &   &   &  1.401 & 220 \bigstrut\\ %& \multirow{3}[6]{*}{400} &
\cline{1-1}\cline{4-4}\cline{12-13}    \#3 &   &   & \multicolumn{1}{c|}{2} &   &   &   &   &   &   &   & 2.281 & 135 \bigstrut\\ %  & 
    \hline
    \multicolumn{13}{c}{} \bigstrut\\ [-1.0em]
    \hline
    \#4 & \multirow{2}[4]{*}{64} & 512 & \multicolumn{1}{c|}{\multirow{2}[4]{*}{8}} & \multirow{2}[4]{*}{64} & Alveo & \multirow{2}[4]{*}{8bit fixed} & \multirow{2}[4]{*}{4157 (46\%)} & \multirow{2}[4]{*}{3148 (78\%)} & \multirow{2}[4]{*}{1284782 (98\%)} & \multirow{2}[4]{*}{661996 (25\%)} & 0.597 & 184 \bigstrut\\ %& \multirow{2}[4]{*}{400}
\cline{1-1}\cline{3-3}\cline{12-13}    \#5 &   & 256 & \multicolumn{1}{c|}{} &   & U55C &   &   &   &   &   &  0.352 & 312 \bigstrut\\
    \hline
    \multicolumn{13}{c}{} \bigstrut\\ [-1.0em]
    \hline
    \#6 & 128 & \multirow{3}[6]{*}{768} & \multicolumn{1}{c|}{\multirow{3}[6]{*}{8}} & \multirow{3}[6]{*}{64} & \multirow{2}[2]{*}{Alveo}  & \multirow{3}[6]{*}{8bit fixed} & \multirow{3}[6]{*}{4157 (46\%)} & \multirow{3}[6]{*}{3148 (78\%)} & \multirow{3}[6]{*}{1284782 (98\%)} & \multirow{3}[6]{*}{661996 (25\%)} & 2 & 314 \bigstrut\\ %& \multirow{3}[6]{*}{400}
\cline{1-1}\cline{2-2}\cline{12-13}    \#7 & 32 &   & \multicolumn{1}{c|}{} &   & \multirow{2}[2]{*}{U55C} &   &   &   &   &   &  0.534 & 285 \bigstrut\\ %& 
\cline{1-1}\cline{2-2}\cline{12-13}    \#8 & 16 &   & \multicolumn{1}{c|}{} &   &   &   &   &   &   &   &  13 & 16 \bigstrut\\ % & 
    \hline
%    \multicolumn{13}{c}{} \bigstrut\\ [-1.0em]
%    \hline
%    \#9 & \multirow{2}[4]{*}{64} & \multirow{2}[4]{*}{768} & \multicolumn{1}{c|}{\multirow{2}[4]{*}{8}} & 32 & Alveo & \multirow{2}[4]{*}{8bit fixed} & 3636 (40\%) & 2636 (65\%) & 746769 (57\%) & 587337 (22\%) & 1.155 & 267 \bigstrut\\ %& \multirow{2}[4]{*}{400}
%\cline{1-1}\cline{5-5}\cline{8-11}\cline{12-13}    \#10 &   &   & \multicolumn{1}{c|}{} & 16 &  U55C &   & 2996 (33\%) & 2380 (59\%) & 607554 (46\%) & 529543 (20\%) &  1.563 & 197 \bigstrut\\ %& 
%    \hline
%    \multicolumn{13}{c}{} \bigstrut\\ [-1.0em]
%    \hline
%    \#11 & \multirow{2}[4]{*}{64} & 768 & \multirow{2}[4]{*}{6} & \multirow{2}[4]{*}{64} & Alveo  & \multirow{2}[4]{*}{8bit fixed} & \multirow{2}[4]{*}{3306 (48\%)} & \multirow{2}[4]{*}{2740 (63\%)} & \multirow{2}[4]{*}{1048022 (88\%)} & \multirow{2}[4]{*}{625983 (26\%)}  & 0.977 & 315 \bigstrut\\ % & \multirow{2}[4]{*}{500}
%\cline{1-1}\cline{3-3}\cline{12-13}    \#12 &   & 512 &   &   & U200 &   &   &   &   &   &  0.604  &  182 \bigstrut\\ 
%    \hline
    \end{tabular}%
    }
  \label{results-overall}%
\end{table}
\vspace{-0.4cm}
\begin{table}[htbp]
\setlength{\arrayrulewidth}{0.05pt}%{0.1mm}
\renewcommand{\arraystretch}{0.9}
  \centering
  \caption{Comparison with Other Acceleration Platforms.}
  \vspace{-0.2cm}
  \resizebox{0.7\columnwidth}{0.9cm}{%1.0\columnwidth}{1.5cm}{
    \begin{tabular}{|c|c|c|c|c|c|c|}
    \hline
    \multirow{2}[2]{*}{\textbf{Platform}} & \multicolumn{1}{|c|}{{Intel E5 }} & \multicolumn{1}{c|}{{NVIDIA V100 }} & \multicolumn{1}{|c}{{Intel Xeon  }} & \multicolumn{1}{|c|}{{NVIDIA P100 }} & \multicolumn{2}{c|}{{\textbf{\textit{FAMOUS}}}} \bigstrut[t]\\
    \multicolumn{1}{|c|}{} & \multicolumn{1}{|c|}{{CPU\cite{luo_calabash_2023}}} & \multicolumn{1}{c|}{{GPU\cite{li_unified_2023}}} & \multicolumn{1}{|c}{{CPU\cite{ye_accelerating_2023}}} & \multicolumn{1}{|c|}{{GPU\cite{ye_accelerating_2023}}} & \multicolumn{2}{c|}{{(Alveo U55C FPGA)}} \bigstrut[b]\\
    \hline
    %\noalign{\hrule height 2pt}
    \textbf{Topologies} & \multicolumn{1}{|c|}{64, 768, 12} & \multicolumn{1}{c|}{64, 512, 4} & \multicolumn{1}{c|}{64, 512, 8} & \multicolumn{1}{c|}{64, 512, 4} & \multicolumn{1}{c|}{64, 768, 8} & \multicolumn{1}{c|}{64, 512, 8} \bigstrut\\
    \hline
    %\textbf{Frequency} & \multirow{2}[2]{*}{2200} & \multirow{2}[2]{*}{1530} & \multirow{2}[2]{*}{2200} & \multirow{2}[2]{*}{1480} & \multicolumn{2}{c|}{\multirow{2}[2]{*}{400}} \bigstrut[t]\\
    %\textbf{(MHz)} &   &   &   &   & \multicolumn{2}{c|}{} \bigstrut[b]\\
    %\hline
    \textbf{GOP} & 0.308 & 0.11 & 0.11 & 0.11 & 0.308 & 0.11 \bigstrut\\
    \hline
    \textbf{Latency} & \multirow{2}[2]{*}{1.1} & \multirow{2}[2]{*}{1.5578} & \multirow{2}[2]{*}{1.96} & \multirow{2}[2]{*}{0.496} & \multirow{2}[2]{*}{0.94} & \multirow{2}[2]{*}{0.597} \bigstrut\\
    \textbf{(ms)} &   &   &   &  &  & \multicolumn{1}{c|}{} \bigstrut[b]\\
    \hline
    \textbf{GOPS} & 280 & 71 & 56 & 221 & 328 & 184 \bigstrut\\
    \hline
    \end{tabular}%
    }
  \label{compare}%
\end{table}%
\vspace{-0.4cm}
\begin{table}[htbp]
\setlength{\arrayrulewidth}{0.05pt}%{0.1mm}
\renewcommand{\arraystretch}{1}
  \centering
  \caption{Comparison with FPGA Accelerators.}
  \vspace{-0.2cm}
  \resizebox{0.65\columnwidth}{3.0cm}{
    \begin{threeparttable}[b]
    \begin{tabular}{|c|c|c|c|c|c|c|}
    \hline
    \multirow{2}[2]{*}{\textbf{Works}} & Calabash & Lu et al. & Ye et al. & Li et al. & Peng et al. & \multirow{2}[2]{*}{\textbf{\textit{FAMOUS}}} \bigstrut[t]\\
      &  \cite{luo_calabash_2023} & \cite{lu_hardware_2020}  &  \cite{ye_accelerating_2023} &  \cite{li_unified_2023} &  \cite{peng_accelerating_2021} &  \bigstrut[b]\\
    \hline
%%    \textbf{Topologies} & 64, 768, 12 & 64, 512, 8 & 64, 512, 4 & 64, 512, 4 & 32, 800, 4 & 64, 768, 8 \bigstrut\\
%%    \hline
    \multirow{2}[2]{*}{\textbf{FPGAs}} & Xilinx  & Xilinx  & Alveo & Xilinx  & Alveo  & Alveo  \bigstrut[t]\\
      & VU9P & VU13P &  U250 & VU37P & U200 & U55C \bigstrut[b]\\
    \hline
    %\textbf{Frequency} & \multirow{2}[2]{*}{243} & \multirow{2}[2]{*}{200} & \multirow{2}[2]{*}{300} & \multirow{2}[2]{*}{200} & \multirow{2}[2]{*}{--} & \multirow{2}[2]{*}{400} \bigstrut[t]\\
    %\textbf{(MHz)} &   &   &   &   &   &  \bigstrut[b]\\
    %\hline
%%    \textbf{Data format} & 16 bit fix & 8 bit fix & 16 bit fix & 8 bit fix & -- & 8 bit fix \bigstrut\\
%%    \hline
    \textbf{Method} & HDL & HDL & HDL & HLS & HLS & HLS \bigstrut\\
    \hline
    \textbf{DSPs} & 4227 & 129 & 4189 & 1260 & 623 & 4157 \bigstrut\\
    \hline
    \textbf{BRAMs} & 640 & 498 & 1781 & 448 & -- & 3148 \bigstrut\\
    \hline
    \textbf{GOPS} & 1288 & 128 & 171 & 72 & 97 & 623 \bigstrut\\ %1350
    \hline
    \textbf{Latency} & \multirow{2}[2]{*}{0.239\tnote{a}} & \multirow{2}[2]{*}{0.8536\tnote{b}} & \multirow{2}[2]{*}{0.642} & \multirow{2}[2]{*}{1.5264} & \multirow{2}[2]{*}{1.706\tnote{c}} & \multirow{2}[2]{*}{0.494} \bigstrut[t]\\
    \textbf{(ms)} &   &   &   &   &   &  \bigstrut[b]\\
    \hline
    \end{tabular}%
    \begin{tablenotes}
        \item [a] Q, K, V matrix computation time ignored.
        \item [b] Time adjusted for 8 attention heads.
        \item [c] Time extracted for attention mechanism from a full transformer.
    \end{tablenotes}
    \end{threeparttable}
    }
  \label{fpga}%
\end{table}%
\vspace{-0.35cm}
%To ensure a fair comparison, we presented the latency specifically for the computation of the attention mechanism, excluding the latency associated with load and store operations for the accelerator, in this table. Our latency is lower and GOPS is higher than all other works except for Calabash\cite{luo_calabash_2023} because it excluded computation time for Q, K, and V calculations. Calabash claims to outperform the Intel E5 CPU with a $2.5 - 6.7\times$ speedup. Therefore, its latency and GOPS could be calculated using the Intel E5 CPU's data from Table~\ref{compare}. In summary, we were able to utilize more DSPs and BRAMs in parallel, even with HLS implementation, achieving the lowest latency for a larger embedding dimension.

%% file: conclusion.tex
\section{Conclusion}\label{conclude} %\& Future Works
\vspace{-0.1cm}
This research presents a flexible FPGA-based accelerator for the multi-head attention layer of transformer neural networks, designed using high-level synthesis. It supports runtime programmability for various topologies without requiring synthesis. Efficient tiling enables large models to fit on-chip while optimizing computation. The accelerator achieves 328 GOPS throughput, outperforming some CPUs and GPUs, with 1.3× lower latency than the fastest state-of-the-art FPGA-based solutions.
\vspace{-0.25cm}